\documentclass[a4paper,fleqn,usenatbib]{mnras}

\usepackage{newtxtext,newtxmath}

\usepackage[T1]{fontenc}
\usepackage{ae,aecompl}

\usepackage{graphicx}	
\usepackage{amsmath}	
\usepackage{amssymb}	
\usepackage{deluxetable}
\usepackage{natbib}
\usepackage{multirow} 
\usepackage{bm} 
\usepackage{threeparttable}
\usepackage{hyperref}
\usepackage{epstopdf}
\usepackage{float}
\usepackage{mathrsfs}

\title[Multiwavelength Analysis of the Variability of the Blazar 3C~273]{Multiwavelength Analysis of the Variability of the Blazar 3C~273}

\author[S. Fernandes]{S. Fernandes$^{1}$,  {V. M. Pati\~no-\'Alvarez$^{2,3}$\thanks{E-mail:
victorp@inaoep.mx, victorm.patinoa@gmail.com (VMP-A)}}, {V. Chavushyan$^{2}$},  {E. M. Schlegel$^{1}$},
\newauthor
and {J. R. Vald\'es$^{2}$} \\
\\
$^{1}$University of Texas at San Antonio, Department of Physics and Astronomy, One UTSA Circle, San Antonio Texas, 78249 TX, USA \\
$^{2}$Instituto Nacional de Astrof\'isica, \'Optica y Electr\'onica, Luis Enrique Erro $\# 1$,
Tonantzintla, Puebla 72840, M\'exico \\
$^{3}$Max-Planck-Institut f\"ur Radioastronomie, Auf dem H\"ugel 69, 53121 Bonn, Germany
}

\date{Accepted XXX. Received YYY; in original form ZZZ}

\pubyear{2020}

\begin{document}
\label{firstpage}
\pagerange{\pageref{firstpage}--\pageref{lastpage}}
\maketitle

\begin{abstract}
We present multiwavelength light curves and polarimetric data of the Flat Spectrum Radio Quasar 3C~273 over 8 years. The wavelength range of our data set extends from radio to gamma-rays. We found that the optical emission in this source is dominated by the accretion disk during the entire time-frame of study. We additionally find that in contrast with the observed behaviour in other blazars, 3C~273 does not show a correlation between the gamma-ray spectral index and the gamma-ray luminosity. Finally, we identified an anti-correlation between the 15~GHz and {\it V}-band light curves for the time-range $JD_{245}=4860 - 5760$, which we speculate is the consequence of the inner part of the accretion disk falling into the black hole, followed by the ejection of a component into the jet.
\end{abstract}

\begin{keywords}
galaxies: active -- galaxies: jets -- gamma-rays: galaxies -- quasars: individual: 3C~273
\end{keywords}




\section{Introduction}
\label{intro}

Historically, AGN have been classified as radio loud (RL) or radio quiet (RQ), meaning whether the ratio of the radio flux density (5 GHz) to the optical flux (B-band) is higher or lower than 10, respectively \citep{Kellermann1989}. However, recent works \citep[e.g.][]{Padovani2016,Padovani2017,Foschini2017} propose to revise these classifications for various reasons. The fact that the difference between RL and RQ AGN is not limited to the properties of their radio emission, but comprises the entire electromagnetic spectrum. The RL AGN emit mainly via non-thermal processes (mostly related to the jet), while the emission from RQ AGN is dominated by thermal processes (related to the accretion disk). On this regard, the main difference on their physical properties is related to the presence or absence of a powerful relativistic jet. Therefore, a new division based on physical properties, instead of observational ones, was proposed: jetted and non-jetted AGN.

Blazars are a sub-class of active galactic nuclei (AGN) where a jet of plasma is closely aligned with our line of sight \citep{Blandford1974, Blandford1978, Blandford1979, Marscher1980, Ghisellini1993, Urry1995}, therefore, classifying them as jetted AGN. These jets can range from being mildly relativistic to ultra relativistic \citep{Marscher2006, Hovatta2009}. Blazars are the dominant population of gamma-ray emitting sources detected by the Fermi Gamma-Ray Space Telescope \citep{Nolan2012}.  Observations of blazars show that they are highly variable across the entire electromagnetic spectrum \citep[e.g. ][]{Urry1996,Ulrich1997,Aharonian2007} and with widely varying time-scales even within the same wavelength band \citep[e.g. ][]{Maraschi1994, Wagner1995}. The spectral energy distribution (SED) has a double hump shape of primarily non-thermal emission extending from the radio to gamma-ray \citep{Marscher1980,Konigl1981,Ghisellini1985,Sambruna1996,Fossati1998, Bottcher2007a, Abdo2010b}. The lower-energy hump is attributed to synchrotron emission from the jet, while the higher-energy hump is attributed to inverse Compton processes in leptonic models and processes such as hadron interactions and pion decay in hadronic models \citep{Bottcher2013}. The nature of the seed photons for the inverse Compton emission is still an active area of debate \citep[e.g. ][]{Maraschi1992,Dermer1992,Sikora1994,Marscher1996, Blazejowski2000, Bottcher2007b, Sikora2009}. \\

There are two sub-classes of blazars: Flat Spectrum Radio Quasars (FSRQ) and BL Lacertae type objects (BL Lacs). FSRQ tend to show broad optical emission lines in addition to a non-thermal continuum, whilst BL Lacs tend to have a featureless non-thermal optical spectrum \citep[e.g.][]{Urry1995, VeronCetty2000}. Blazars are also known to have highly variable linear polarisation from radio to optical \citep[e.g.][]{Villforth2010,Ikejiri2011,Falomo2014} that ranges from less than 2\% \citep[e.g.][]{Impey1989} to over 30\% as seen in other blazars \citep[e.g.][]{Zhang2015,PatinoAlvarez2018}. However, these observational characteristics can change depending on the activity state of the object \citep{Ghisellini2011}

3C~273, a source at z = 0.158 was the first identified quasar, although it was initially classified as a star \citep{Iriarte1957}. The first identification as a quasar of 3C~273 was in the 3rd Cambridge Catalogue \citep{Edge1959}. The first studies focused on identifying the components observed in the optical, the position of the object \citep{Hazard1963}, and the redshift from spectral lines \citep{Schmidt1963}. 3C~273 is highly luminous, especially in the optical \citep{Courvoisier1998}. It is in the blazar sub-class of FSRQ \citep{Penston1970}. \\

Since 3C~273 was the first discovered quasar, it has been extensively studied across all wavelengths. There is considerable literature on the highly variable nature of 3C~273 from radio to gamma-rays, therefore, here we provide only some of the most relevant works for our particular study. There is an excellent review on the literature and state of knowledge of 3C~273 up to 1998 by \citet{Courvoisier1998}. Morphological studies have been done in the radio using Very Long Baseline Interferometry (VLBI) imaging \citep{Attridge2005}, where the authors found two strongly polarised components within a milliarcsecond of the core. The jet has been studied in X-rays \citep{Sambruna2001}, where it was found that the spectral energy distributions of four selected regions can best be explained by the inverse Compton scattering. \cite{Jester2005} showed that the infrared-ultraviolet spectral index presents an important flattening; as well as a slow decrease of the maximum energy in the jet. This implies particle re-acceleration acting along the entire jet. \cite{Savolainen2006}, using the Very Long Baseline Array (VLBA), resolved the emission components in the pc-scale jet of 3C~273, and showed how the different components moved along the jet during the observation campaign of 2003. A more recent view on the pc-scale jet of 3C~273 is presented in \cite{Lisakov2017}, using data from the Boston University VLBA monitoring programme for blazars, analysed the kinematics of multiple newborn components, and concluded that the gamma-ray emission zone is close to the jet apex, 2-7 pc upstream from the observed 7mm core. \\

The radio variability was one of the earliest focuses of multiwavelength studies \citep[e.g.][]{Aller1985, Terasranta1992, Reich1993, Mantovani2000}. Optical and colour-index studies have been carried out along with estimations of the central black hole mass \citep[e.g.][]{Kaspi2000, Paltani2005, Dai2009, FanJH2009, FanJH2014}. High energy variability has been studied in the context of gamma-ray flaring and the origin of X-ray emission \citep[e.g.][]{Collmar2000, Kataoka2002, Abdo2010d}. The correlation between the IR and X-ray has been studied by \cite{McHardy2007}. Optical to X-ray and gamma-ray studies have been carried out in part to study the high energy end of the spectral energy distribution \citep[SED, e.g.][]{Courvoisier2003, Pacciani2009, Kalita2015}. To identify radio and gamma-ray emitting regions, \cite{LiuHT2015} studied radio variability along with the line variability of H$\alpha$, H$\beta$, H$\gamma$, Ly$\alpha$, and CIV. Full multiwavelength coverage, from the radio to gamma-rays, has been carried out since the first gamma-ray observatories \citep[e.g.][]{Courvoisier1987, Turler1999, Soldi2008, Volvach2013, Chidiac2016}. \\

The radio and optical properties of 3C~273 have been observed with increasing levels of resolution since the first observations in 1963 \citep[e.g.][]{Davis1985,Flatters1985, Roser1991,Valtaoja1991, Courvoisier2003, Marscher2004, Stawarz2004}; as well as superluminal jet components \citep[e.g.][]{Belokon1991, LiuWP2015}. \\

\citet{Kaspi2000} carried out reverberation mapping based on 39 observations made over seven years (1991-1998); they reported a delay of $342^{+145}_{-89}$ days between the H$\beta$ emission line and the continuum emission; which yields a black hole mass of $(5.5^{+0.89}_{-0.79}) \times 10^8$ M$_{\odot}$. On the other hand, \cite{Zhang2019} also performed reverberation mapping for 3C 273, using 283 observations over a period of 10 years (2008-2018); they report a time delay of 146.8$^{+8.3}_{-12.1}$ days, which applied with a virial factor $f_{BLR}$ of 1.3, results in a black hole mass of $(4.1^{+0.3}_{-0.4}) \times 10^8$ M$_{\odot}$. \\

In this paper we present a multiwavelength and polarimetry analysis of 3C~273 from 2008 to 2015. We focus on correlated behaviour between wavelengths and the physical implications of such correlations. \\


\begin{figure*} 
\includegraphics[width=1.0\textwidth]{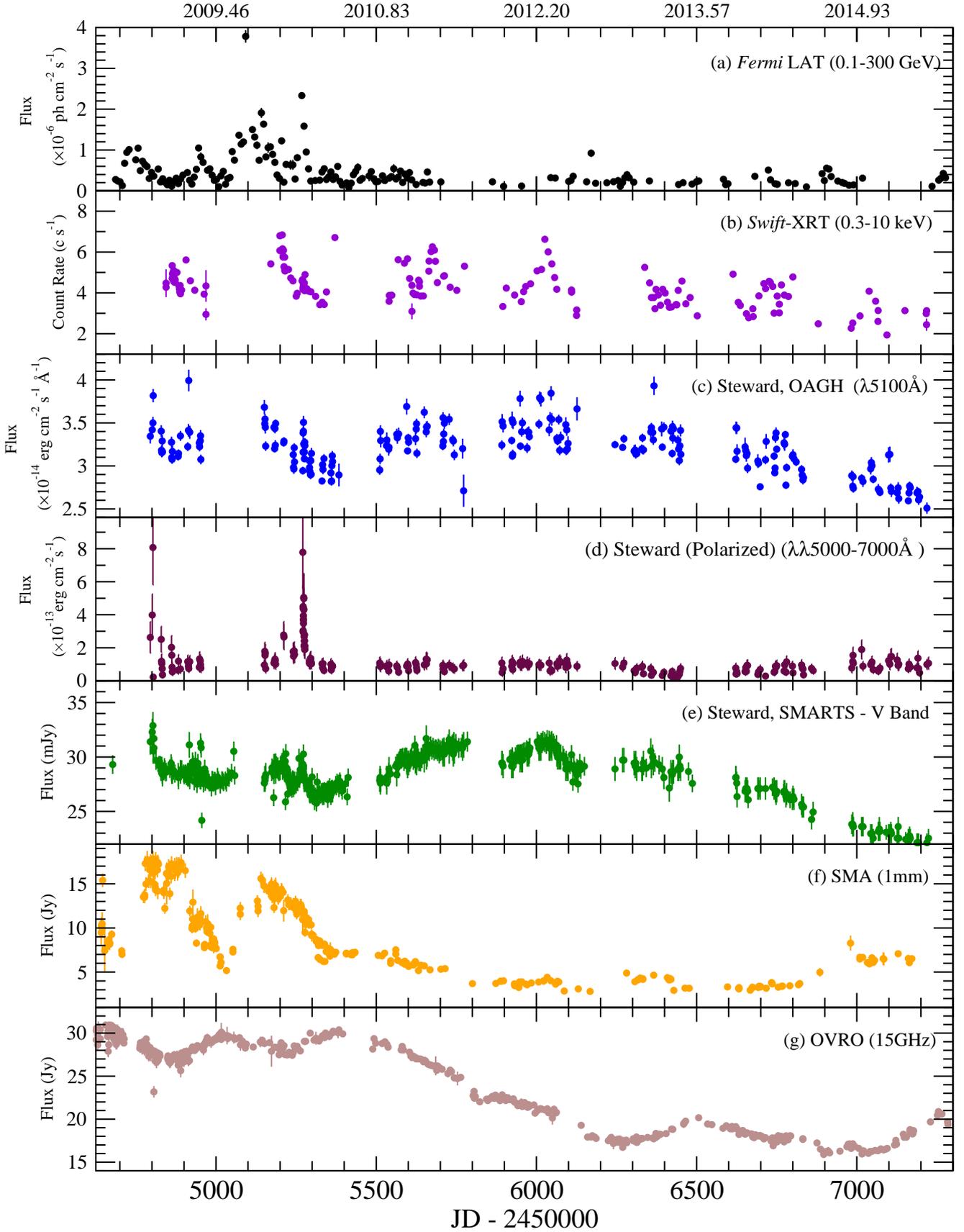}
\caption{Multiwavelength light curves for 3C 273. a) Gamma-rays, b) X-rays, c) optical spectral continuum, d) polarised flux, e) V-band, f) J-band, g) 1~mm, and h) 15~GHz. Origin of the data, as well as spectral band are labeled in each panel.} 
\label{mw}
\end{figure*}

\section{Observations and Data Reduction} 
\label{obsreduc}

The multiwavelength data used in this work were obtained from many different sources. Gamma-ray data were retrieved from the public archive of the \textit{Fermi} Large Area Telescope (\textit{Fermi-LAT}) \citep{Atwood2009}. X-ray data were taken from the \textit{Swift}-XRT Monitoring of \textit{Fermi}-LAT Sources of Interest webpage\footnote{http://www.swift.psu.edu/monitoring/} \citep{Stroh2013}. Optical V-band, spectroscopy, and spectropolarimetry data were obtained from the Ground-based Observational Support of the \textit{Fermi Gamma-ray Space Telescope} at the University of Arizona\footnote{http://james.as.arizona.edu/$\sim$psmith/Fermi/} \citep[Steward Observatory,][]{Smith2009}. Optical V-band data was obtained from the Yale \textit{Fermi}/SMARTS program \citep{Bonning2012}. Optical spectra were also observed at the Observatorio Astrof\'isico Guillermo Haro (OAGH) \citep{PatinoAlvarez2013}. One millimetre data were taken from the Sub-Millimeter Array database \citep[SMA,][]{Gurwell2007}. 15~GHz radio continuum observations were taken from the public database of the Owens Valley Radio Observatory (OVRO) 40 m Telescope Blazar Monitoring Program \citep{Richards2011}. Details on the different datasets are summarised in Table \ref{datadesc}.

We processed the Fermi-LAT gamma-ray data in the energy range from 0.1 to 300 GeV using the \texttt{Fermitools version 1.0.2}. We analysed a region of interest (ROI) of 15$^{\circ}$ in radius around the position of 3C~273, and included all sources within, which were extracted from the 4FGL catalogue \citep{4FGL}. All of the normalisation factors and spectral indices of sources within 5$^{\circ}$ were kept free; while sources farther than 5$^{\circ}$ only had their normalisation parameter free. We also used up-to-date diffuse and isotropic background models along with the current set of instrument response functions. We binned the data in 7-days intervals to increase the signal to noise ratio, and used only those data points with a Test Statistic (TS) value greater than 25 (significant detection). For the data summary see Table \ref{datadesc} and for the light curve (using the recommended log-parabola spectral model) see panel (a) of Figure~\ref{mw}. 

We constructed the X-ray light curve in the energy range from 0.3 to 10 keV by retrieving reduced X-ray data from the \textit{Swift}-XRT Monitoring of \textit{Fermi}-LAT Sources of Interest \citep[For full details of the analysis techniques for the data set see][]{Stroh2013}. For the data summary see Table \ref{datadesc} and for the light curve see panel (b) of Figure~\ref{mw}. \\

We obtained V-band data from the SMARTS project \citep{Bonning2012} and the Steward Observatory Monitoring Program \citep{Smith2009}. There is a systematic difference between the magnitudes of the two data sets for this source, due to the use of very faint stars for calibration in the case of the SMARTS data (Isler 2015, private communication). Therefore, we decided to apply a correction to the SMARTS magnitudes. This correction factor was calculated by obtaining the magnitude differences in data from the same observing night for both databases. A visual inspection of the distribution of these differences shows a gaussian pattern, with a mean and standard deviation of $0.08 \pm 0.03$. In this case the correction factor is added to the SMARTS magnitude, while the standard deviation is added quadratically to the uncertainty in the magnitude. We estimated the error in the flux using the error in the measured magnitude, the error in the absolute calibration of the photometry, and standard error propagation. We constructed the light curve of the V-band in panel (e) of Figure~\ref{mw} combining the SMARTS and Steward Observatory data sets. For the data summary see Table \ref{datadesc}. \\

Fully reduced 1~mm data were retrieved from the SMA Observer Centre website\footnote{http://sma1.sma.hawaii.edu/callist/callist.html} \citep{Gurwell2007}. For the data summary see Table \ref{datadesc} and for the light curve see panel (g) of Figure~\ref{mw}. 

We obtained fully reduced data from the OVRO blazar monitoring program \footnote{\url{http://www.astro.caltech.edu/ovroblazars/data/data.php}}. The OVRO 40~m telescope blazar monitoring program makes regular 15~GHz observations of many objects including 3C~273. The telescope uses off-axis dual-beam optics and a cryogenic high electron mobility transistor (HEMT) low-noise amplifier with a 15~GHz centre frequency and 3~GHz bandwidth. Full details of the data reduction and instrument are given in \cite{Richards2011}. There is a systematic uncertainty in the flux-density of about 5 \%. For the data summary see Table \ref{datadesc} and for the light curve see panel (h) of Figure~\ref{mw}.

\begin{table*}
\centering
 \begin{minipage}{0.9\textwidth}
 \caption{Description of the data sets: 3C~273}
   \begin{tabular}{ c c c c c }
  \hline
  \textbf{Data} & \textbf{Date Range} & \textbf{Date Range} & \textbf{Number of} & \textbf{Origin of} \\
  & \textbf{($JD_{245}$)} & (Calendar) & \textbf{Observations} & \textbf{Data} \\
  \hline
  \hline
  Gamma Rays & \multirow{2}{*}{4686.656 - 7059.656} & \multirow{2}{*}{2008 August 8 - 2015 February 6} & \multirow{2}{*}{191 (bins)} & \multirow{2}{*}{\textit{Fermi} LAT} \\
  (0.1 - 300 GeV) & & & \\
  \hline
  Soft X-ray & \multirow{2}{*}{3418.2561 - 7217.8396} & \multirow{2}{*}{2005 February 16 - 2015 July 14}  & \multirow{2}{*}{210}  &  \multirow{2}{*}{\textit{Swift} XRT} \\
 (0.3-10 keV) & & & \\
  \hline
  Optical Continuum & \multirow{2}{*}{4795.0177 - 7223.6617} & \multirow{2}{*}{2008 November 24 -  2015 July 20} & \multirow{2}{*}{236} & OAGH \\
  ($\lambda$5100 \AA) & & & & Steward Observatory\\
  \hline
  V Band & \multirow{2}{*}{4677.4987 - 7223.6695} & \multirow{2}{*}{2008 July 29 - 2015 July 20} & \multirow{2}{*}{553} & SMARTS \\
 ($\lambda_{eff}$ = 5510 \AA) & & & & Steward Observatory \\
  \hline
  Submillimetre & \multirow{2}{*}{3413.6701 - 7171.6743} & \multirow{2}{*}{2005 February 12 - 2015 29 May} & \multirow{2}{*}{492} & \multirow{2}{*}{SMA} \\
  (1 mm) & & & \\
  \hline
  Radio & \multirow{2}{*}{4473.9830 - 7431.5348} & \multirow{2}{*}{2008 January 8 - 2016 February 13} & \multirow{2}{*}{602} & \multirow{2}{*}{OVRO} \\
 (15 GHz) & & & \\
  \hline
  Polarimetry & \multirow{2}{*}{4795.02 - 7223.66} & \multirow{2}{*}{2008 November 24 - 2015 July 20} & \multirow{2}{*}{290} & \multirow{2}{*}{Steward Observatory} \\
($\lambda\lambda$5000-7000 \AA) & & & \\
  \hline
  \end{tabular}
  \label{datadesc}
  \end{minipage}
  \end{table*}

\subsection{Optical Spectra}

\begin{figure}
\includegraphics[width=0.5\textwidth]{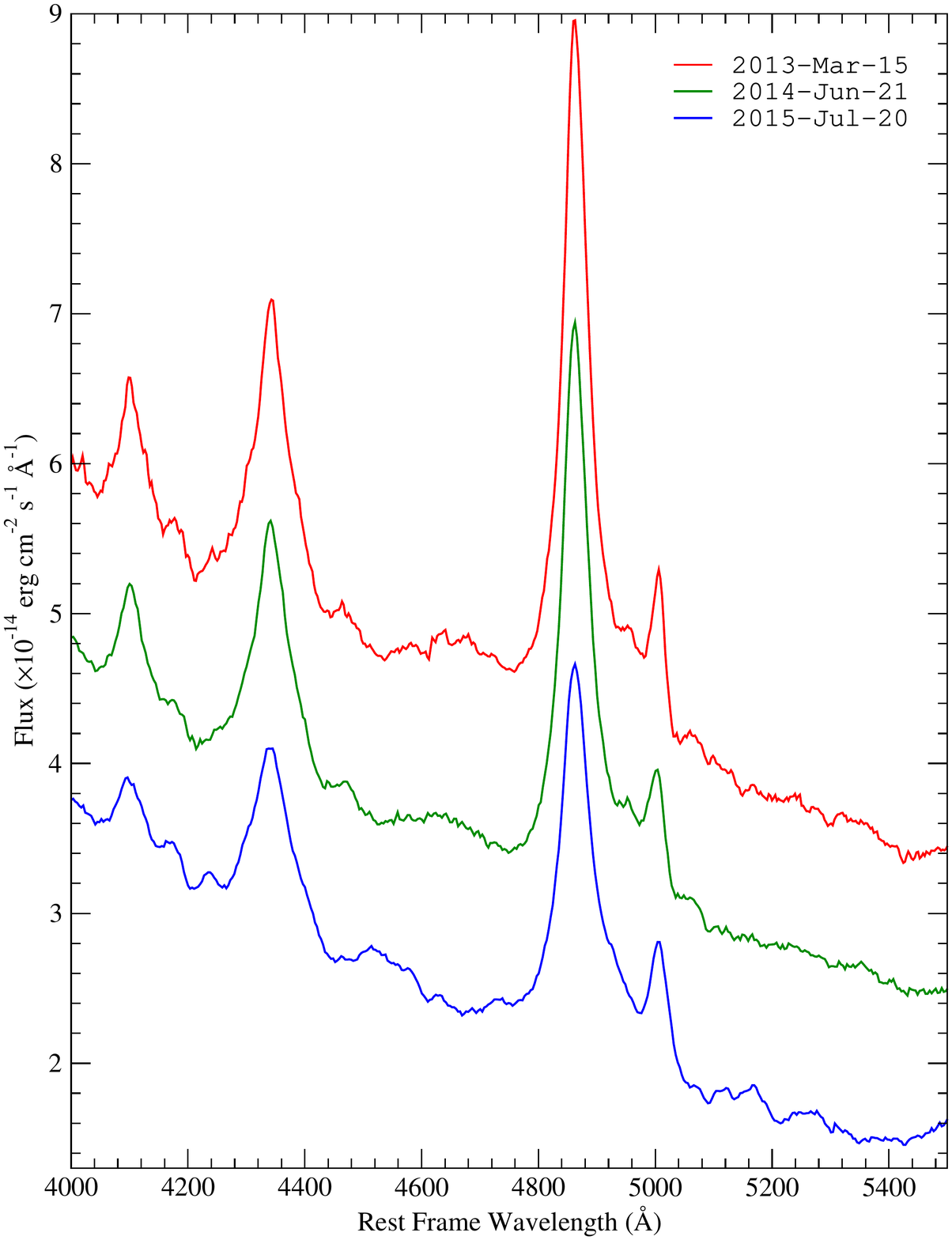}
\caption{Spectra of 3C~273 at different activity states. The red spectrum corresponds to the date of the highest continuum flux observed, the blue spectrum 
corresponds to the lowest continuum flux observed, and the green spectrum corresponds to a medium point in continuum flux between the red and blue spectra.}
\label{spectvar}
\end{figure}

The optical spectra used in this work come from two different sources. Nine spectra were taken at The Observatorio Astrof\'isico Guillermo Haro (OAGH), while 227 spectra were taken at the Steward Observatory. The Steward Observatory spectra were taken in the framework of the Ground-based Observational Support of the Fermi Gamma-ray Space Telescope at the University of Arizona monitoring program\footnote{http://james.as.arizona.edu/$\sim$psmith/Fermi/}. Details on the observational setup and reduction process are presented in \cite{Smith2009}. In this work we only use spectra that have been re-calibrated against the V-band magnitude. Examples of spectra at different activity states are shown in Figure~\ref{spectvar}.

Within the framework of a spectrophotometric monitoring programme of bright gamma-ray sources at OAGH in Sonora, M\'exico \citep{PatinoAlvarez2013}, we carried out spectroscopic observations of the blazar 3C~273 using the Boller \& Chivens long-slit spectrograph on the 2.1~m telescope at OAGH. The spectra were obtained using a slit width of 2.5 arcsec, and during photometric weather conditions. The spectral resolution was $R=15$ \AA\ (FWHM). The wavelength range for the spectra is 3800 to 7100 \AA. The S/N ratio is $>40$ in the continuum near H$\beta$ for all spectra. To enable a wavelength calibration, He-Ar lamp spectra were taken after each object exposure. At least two spectrophotometric standard stars were observed each night, to enable flux calibration. Table \ref{obslog} shows the observation log for the OAGH spectra.

The spectrophotometric data reduction was carried out using the \textit{IRAF} Package\footnote{\url{http://iraf.noao.edu/}}. The reduction process included bias and flat-field corrections, cosmic ray removal, 2-D wavelength calibration, sky spectrum subtraction, and flux calibration. The 1-D spectra was subtracted taking an aperture of 6 arcsec around the peak of the spectrum profile.

\begin{table*}
\centering
\begin{minipage}{190mm}
\caption{Observational log of the spectra taken at OAGH.}
\begin{tabular}{ccccccccc}
\hline
UT Date & Observatory & Tel. + Equipment & Aperture & Focus & Grating & Resolution ($\AA$) & Seeing & Exposure Time (s) \\
\hline
\hline
2013 Feb 08 & OAGH & 2.1m + B \& C & 2.5"$\times$6.0" & Cassegrain & 150 l/mm & 15.4 & 2.3" & 3$\times$1800 \\
2013 Mar 15 & OAGH & 2.1m + B \& C & 2.5"$\times$6.0" & Cassegrain & 150 l/mm & 15.4 & 2.6" & 2$\times$1800 \\ 
2013 Apr 08 & OAGH & 2.1m + B \& C & 2.5"$\times$6.0" & Cassegrain & 150 l/mm & 15.4 & 3.3" & 2$\times$1800 \\ 
2013 May 12 & OAGH & 2.1m + B \& C & 2.5"$\times$6.0" & Cassegrain & 150 l/mm & 15.4 & 2.5" & 4$\times$1200 \\ 
2013 Jun 07 & OAGH & 2.1m + B \& C & 2.5"$\times$6.0" & Cassegrain & 150 l/mm & 15.4 & 1.9" & 3$\times$1200 \\ 
2014 Feb 28 & OAGH & 2.1m + B \& C & 2.5"$\times$6.0" & Cassegrain & 150 l/mm & 15.4 & 2.0" & 3$\times$1800 \\
2014 Mar 28 & OAGH & 2.1m + B \& C & 2.5"$\times$6.0" & Cassegrain & 150 l/mm & 15.4 & 2.3" & 3$\times$1800 \\
2014 Apr 29 & OAGH & 2.1m + B \& C & 2.5"$\times$6.0" & Cassegrain & 150 l/mm & 15.4 & 2.0" & 3$\times$1800 \\
2014 Jun 02 & OAGH & 2.1m + B \& C & 2.5"$\times$6.0" & Cassegrain & 150 l/mm & 15.4 & 2.0" & 3$\times$1800 \\
\hline
\end{tabular}
\label{obslog}
\end{minipage}
\end{table*}

\subsection{Polarimetry Data}

We obtained the wavelength calibrated, normalised Stokes Q and U spectra from the Steward Observatory monitoring program database\footnote{http://james.as.arizona.edu/$\sim$psmith/Fermi/}. From these spectra we are able to estimate the degree of polarisation ($P$) and the position angle of polarisation ({\it PA}). A full description of the data analysis is given in \cite{PatinoAlvarez2018} and a full explanation of the Steward Observatory observations and data reduction is given in \citet{Smith2009}. 

\begin{figure*}
\centering
\includegraphics[width=1.0\textwidth]{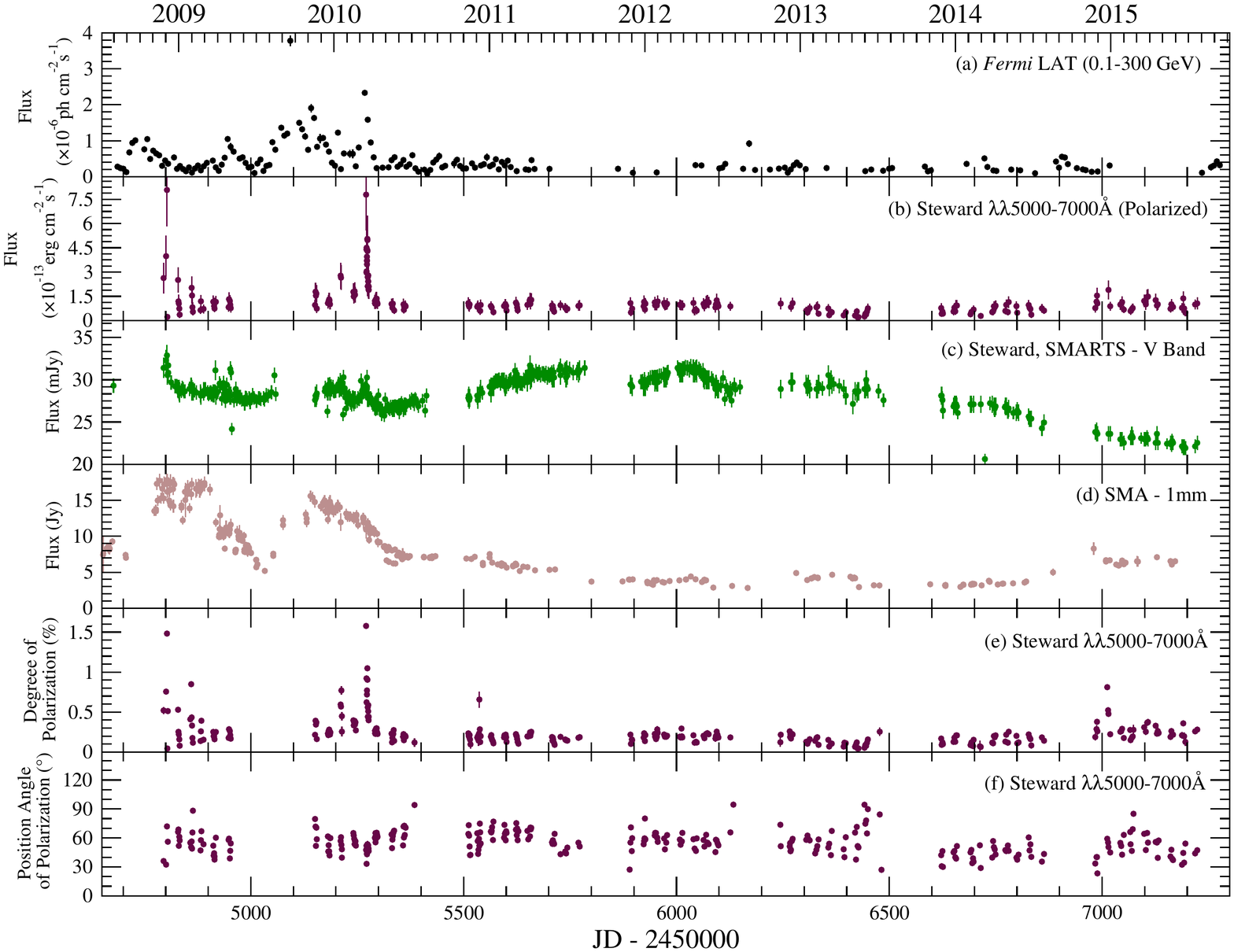}
\caption{Comparison of the light curves in: a) gamma-rays, b) polarised flux, c) V-band, d) 1~mm, e) degree of polarisation ($P$) and f) position angle of polarisation ($PA$).}
\label{polfig}
\end{figure*}

We use the spectral range from $\lambda$5000\AA\ to $\lambda$7000\AA\ for our analysis to avoid noisy edges of the spectra. We made no correction to Galactic interstellar extinction or polarisation due to the high Galactic latitude of 3C~273. To determine if there was a wavelength dependence of the degree of polarisation, we fit a first order polynomial to the degree of polarisation spectra from $\lambda$5000\AA\ to $\lambda$7000\AA. We found no statistically significant gradients in any spectra from the fits and concluded that a single value could characterise the degree of polarisation and position angle of polarisation for this wavelength range for any given observation. For a description of the calculation of polarimetric quantities and errors see \cite{PatinoAlvarez2018}. For the data summary see Table \ref{datadesc} and for the light curves see Figure~\ref{polfig}.\\

\section{Spectral Analysis} 
\label{specanaly}

To obtain the emission line and continuum flux, the spectra from both, the Steward Observatory and OAGH, were first transformed to rest-frame wavelength, while applying a cosmological correction to the monochromatic flux of the form $(1 + z)^3$. Then, the featureless continuum is approximated by a power-law function by applying a least-squares minimisation using the MPFIT Package \citep{Markwardt2009}, and then the continuum is subtracted from the spectrum.

The optical Fe II emission is fitted using the procedure explained in \cite{Kovacevic2010} and \cite{Shapovalova2012}. The fitting is done through a webpage applet\footnote{\url{http://servo.aob.rs/FeII\_AGN/}} where an ASCII file containing the spectrum can be uploaded. The output will be the fitting of the Fe II emission divided by its different emission line groups (F, S, G, P, I Zw 1). The algorithm assumes that all Fe II lines within the considered range originate from the same AGN emission region. This implies the same physical and kinematical conditions in the region in which these lines arise, i.e. the same width for all Fe II lines.\\

\begin{figure}
\includegraphics[width=0.48\textwidth]{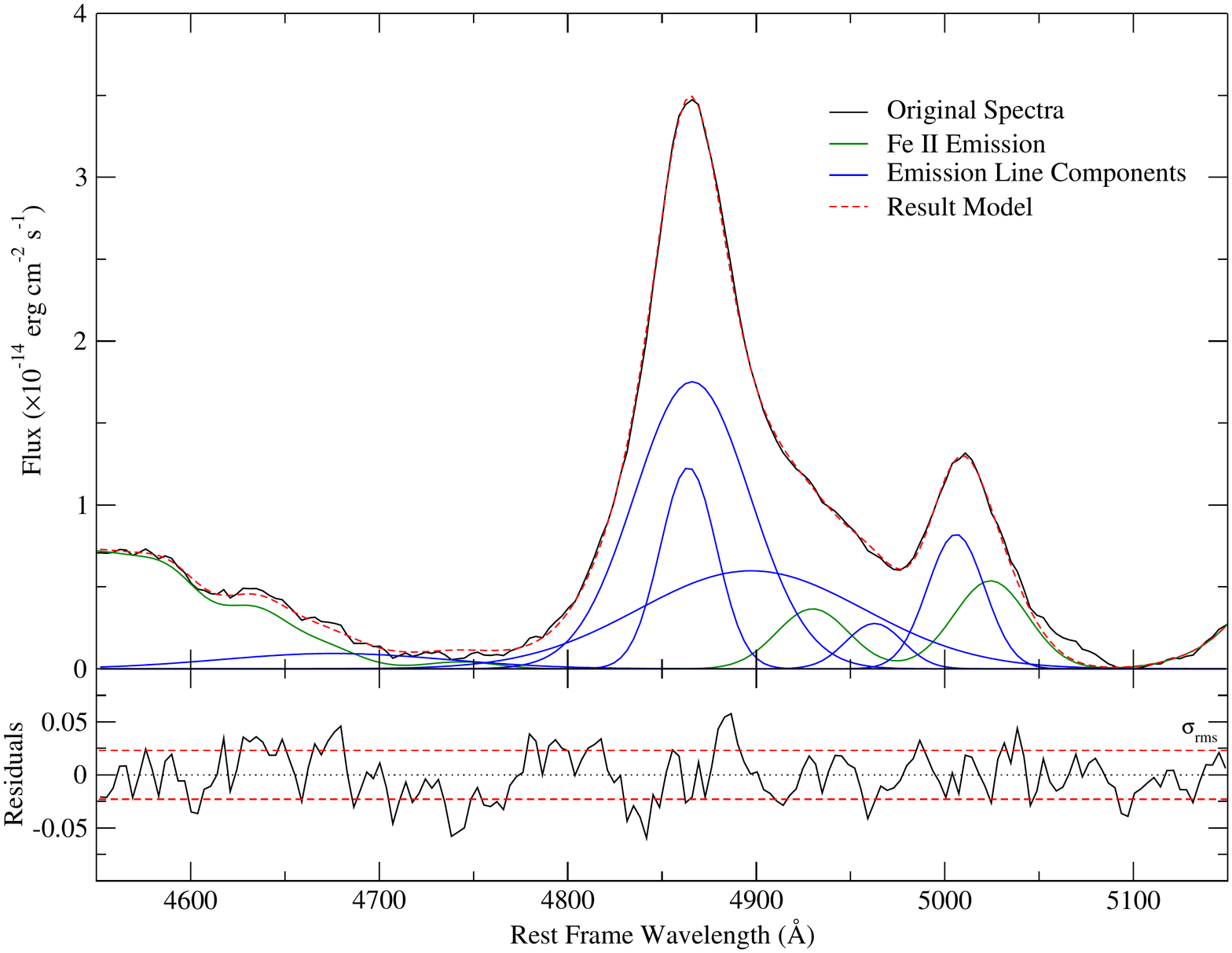}
\caption{Top: Rest frame spectrum of 3C~273, showing the Gaussian decomposition and iron fitting for the spectral region of the H$\beta$ emission line. The black solid line represents the observed spectrum, the blue solid lines represent the Gaussian emission line components, the green solid line represents the Fe II emission, and the red dashed line represents the total model. Bottom: Residuals of the fitting shown in the top panel. The red dashed lines represent the $\sigma_{rms}$ of the residuals. The flux axis is in the same units in both panels.}
\label{irondecomp}
\end{figure}

After the continuum and Fe II subtraction of the spectra, we performed a Gaussian decomposition to all the emission lines near the H$\beta$ line, as explained in \cite{PatinoAlvarez2016}. We assumed that a single value of FWHM is adequate to fit the narrow components of the H$\beta$ and [O III] lines ([O III] $\lambda$5007\AA\ FWHM was adopted); the broad component of H$\beta$ was fitted using two independent Gaussians; the He II $\lambda$4686\AA\ emission line was fitted with a single Gaussian. The theoretical flux ratio between the [O III] emission lines was held fixed during the fitting. An example of the iron-line fitting and Gaussian decomposition can be seen in Figure~\ref{irondecomp}.\\

\subsection{Flux Recalibration by the [OIII] Emission Line and Flux Measurements}
\label{fluxrecal}

To homogenise the spectral measurements made at the Steward Observatory and at OAGH, we performed a flux recalibration of all the spectra. We wanted to reduce the effects of different observational conditions between the two observatories. To recalibrate the spectra we multiplied the spectral flux by a recalibration factor. We calculated this recalibration factor by assuming that the narrow lines in AGN do not vary over short periods of time. This is supported by the large size of the narrow line region (NLR) in these objects \citep[e.g.][]{Bennert2002}. A region of large size will respond more slowly to ionising radiation leading to slow variability time-scale of narrow line emission. This is observed in Seyfert galaxies where the extent of the NLR leads to light travel times of hundreds to thousands of years and recombination time-scales on the order of hundreds of years \citep{Urry1995, Kassebaum1997}. Thus, the narrow emission lines are assumed constant over our time-frame. Therefore, we used the [O III] $\lambda$5007\AA\ narrow emission line as a reference point to calculate the recalibration factor. \\

Given that the spectra taken at the Steward Observatory are re-calibrated by V-band photometry, we decided to use the measurements of these spectra to calculate the average flux of the [O III] $\lambda$5007 \AA. The average flux of the [O III] $\lambda$5007 \AA\ we obtained was $2.816\times10^{-13}$ erg s$^{-1}$ cm$^{-2}$. Then, the calibration factor is calculated by dividing the measured flux by the average [O III] $\lambda$5007 \AA\ flux. Then this factor is applied to all of the spectra.\\

Once all the spectra were recalibrated to have the same flux in the [O III] $\lambda$5007 \AA\ emission line, we proceeded to measure the 
$\lambda$5100 \AA\ optical continuum emission, as well as the H$\beta$ emission line. \\

\subsection{Flux Measurements} \label{fluxmeas}

\begin{figure*}
\includegraphics[width=1.0\textwidth]{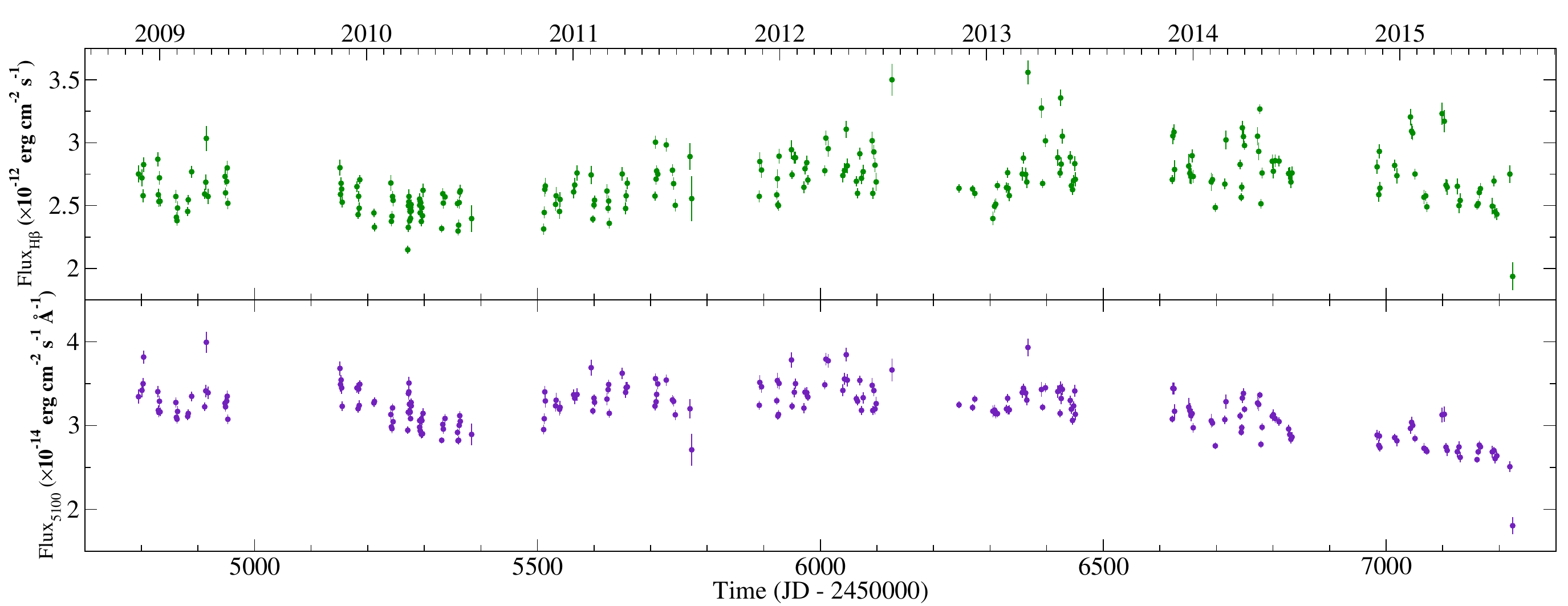}
\caption{Spectroscopic light curves after recalibration by [O III]. Top: H$\beta$ emission line light curve. Bottom: $\lambda$ 5100\AA\ optical continuum light curve.}
\label{linecont}
\end{figure*}

The flux of the H$\beta$ emission line was calculated by adding the fluxes of 3 Gaussian components (one narrow and two broad) that make up the H$\beta$ profile in the spectra used in this work. The uncertainty on the emission-line flux measurement involves two components, the first one is calculated using the formula from \cite{Tresse1999}. This formula includes systematic errors, as it takes into account the characteristics of the spectra, like the S/N ratio, dispersion, intensity of the line compared to the continuum, etc. are all factors that contribute to the uncertainty. The second source of error is the one introduced by the Fe II subtraction, which was calculated as in \cite{LeonTavares2013}. The continuum flux at $\lambda$5100 \AA\ was measured on the spectra after subtracting the Fe II emission, as the mean in the wavelength range 5050-5150 \AA. The uncertainty reported is the RMS over the same wavelength range. The re-calibrated H$\beta$ emission line and $\lambda$5100 \AA\ continuum light curves are shown in Figure~\ref{linecont}.\\

\section{Analysis of the Gamma-Ray Emission} 
\label{gamanaly}

We used two different models to produce light curves of the gamma-ray flux and the spectral index. The model functions are shown in Equations \ref{log-parabola} and \ref{power-law}. This is necessary to accurately fit the parameters for each light curve. We used gamma-rays data from the Fermi Large Area Telescope (LAT). We used Pass 8 data, which was reduced and analysed using Fermitools version 1.0.2, along with the latest diffuse model files and the current set of recommended Instrument Response Functions. We obtained gamma-ray fluxes by maximum likelihood, where the initial model is built with sources within 15 degrees of 3C 273, obtained from the 4th Fermi-LAT Catalogue \citep[4FGL,][]{4FGL}. The spectral parameters for all the sources within 5 degrees were left free to vary, while for sources farther than 5 degrees, only the normalizations were left as free parameters. We only use data from time bins where the Test Statistic (TS) is higher or equal to 25 (detection significance of 5$\sigma$). For the light curve shown in Panel (a) Figure~\ref{mw}, we adopted time bins of 7 days, to ensure sufficient S/N ratio per bin, and the spectrum was modelled with a log-parabola spectral shape (recommended model for 3C 273 in the 4FGL catalogue), as shown in Equation \ref{log-parabola}.

\begin{equation}
\frac{dN}{dE} = N_0 \Bigg( \frac{E}{E_{b}} \Bigg) ^{-(\alpha + \beta log(E/E_0))},
\label{log-parabola}
\end{equation}

where $N_0$ is the normalisation parameter, $E_b$ is the energy scale, $\alpha$ and $\beta$ are indices that describe the curvature of the spectrum.

\subsection{Gamma-Ray Spectral Index} \label{grayspecindex}

In the light of previous studies that postulate a possible correlation between the gamma-ray luminosity and the gamma-ray spectral index \citep[e.g.][]{Ghisellini2009,Foschini2010,Brown2011,FanXL2012,FanJH2012,PatinoAlvarez2018}, we decided to investigate changes on the spectral shape of the high energy emission, by computing fluxes in weekly time bins and obtaining their respective spectral index. The gamma-ray spectral index ($\gamma$) is well-defined only in the context of a power-law spectral model (see Equation \ref{power-law}). Therefore, we computed the gamma-ray fluxes using a power-law spectral model for our source of interest.

\begin{equation}
\frac{dN}{dE} = N_0 \Bigg( \frac{E}{E_0} \Bigg) ^{-\gamma}
\label{power-law}
\end{equation}

\noindent where $N_0$ is the normalisation parameter, $E_b$ is the energy scale, and $\gamma$ is the spectral index. 

Since the power-law spectral model is not the one recommended in the 4FGL for 3C~273, we have to test for each of the bins, if the power-law model accurately describes the gamma-ray spectrum; otherwise, we might be utilizing gamma-ray spectral indices that do not have physical meaning (making any kind of analysis pointless). If the data is indeed well-represented with a power-law model, then the index $\beta$ (Equation~\ref{log-parabola}) should be consistent with zero, in which case, we end up with the power-law spectral model (Equation~\ref{power-law}).\\

Therefore, we propose to use the results from the log-parabola model to discern which time bins have an spectrum well-represented by a power-law model, based on the value of the index $\beta$ mentioned above. By taking the time bins in which the $\beta$ index is consistent with zero within 1-$\sigma$, we obtain a total of 52 time bins, of which, 49 have a significant detection when the power-law model is applied.\\

\begin{figure*}
\includegraphics[width=1.0\textwidth]{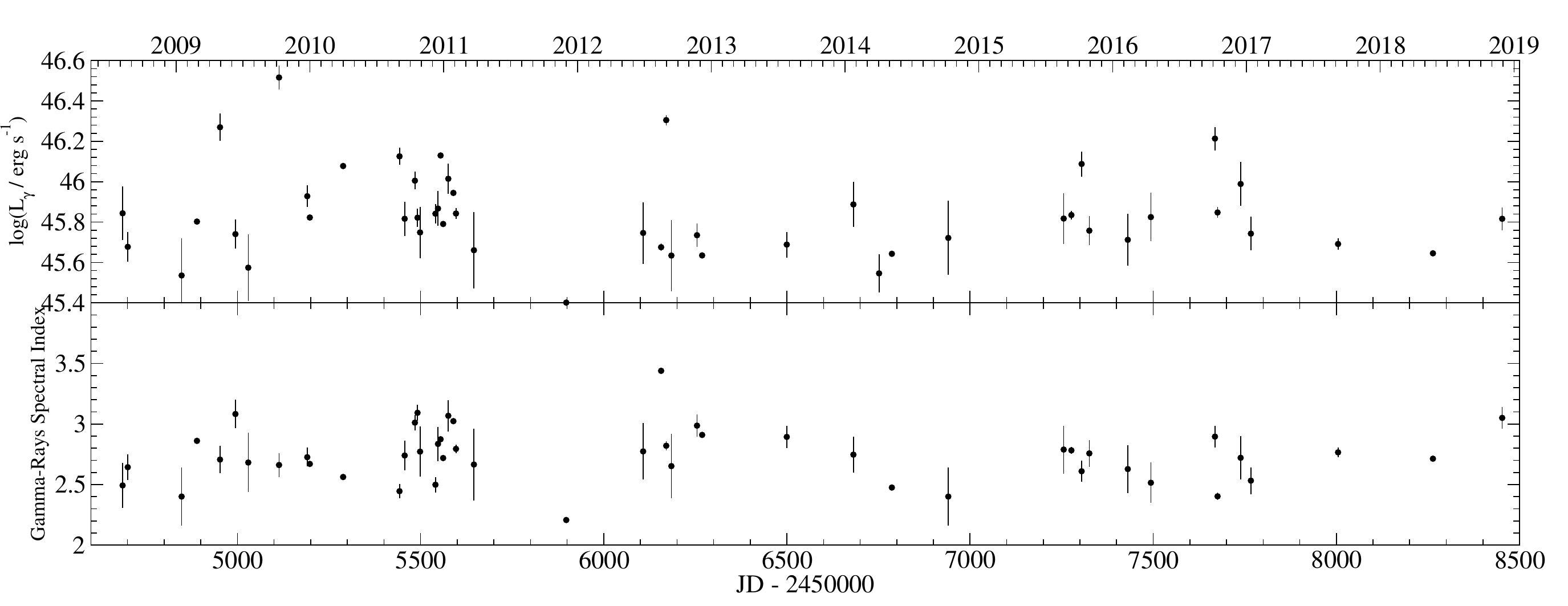}
\caption{Top: Gamma-ray luminosity (power-law model) light curve. Bottom: Gamma-ray spectral index light curve.}
\label{gamind}
\end{figure*}

\begin{figure}
\includegraphics[width=0.48\textwidth]{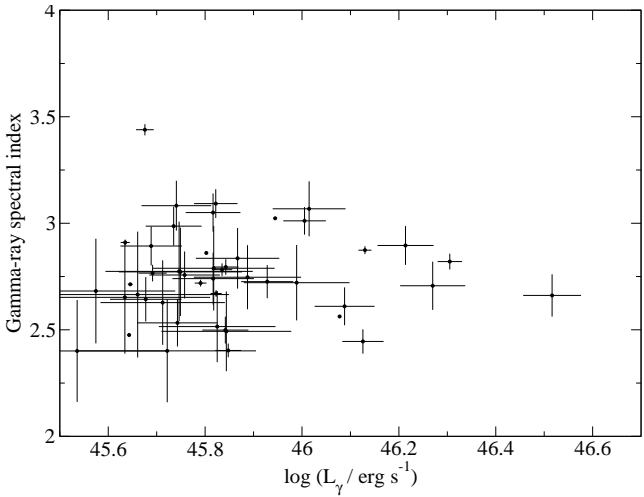}
\caption{Relationship between the gamma-ray luminosity and gamma-ray spectral index. }
\label{gamindlum}
\end{figure}

Next, we proceed to calculate the gamma-ray luminosity following equations 1 and 2 from \cite{Ghisellini2009}. The light curves of the gamma-ray luminosity and spectral index are as shown in Figure~\ref{gamind}. We show the relationship between the gamma-ray luminosity and the gamma-ray spectral index in Figure~\ref{gamindlum}. We decided to show the luminosity light curve instead of the flux, since it will make it easier to compare with the luminosity vs. spectral index plot. To determine if there is a correlation between the gamma-ray luminosity and the spectral index in 3C~273, we performed a Spearman Rank correlation test, which yielded a correlation coefficient ($\rho$) of 0.09, and a p-value (probability of obtaining the correlation by chance) of 0.54, indicating that no correlation was found. This result differs from those found for other blazars in the aforementioned papers \citep{Ghisellini2009,Foschini2010,Brown2011,FanXL2012,FanJH2012,PatinoAlvarez2018}, however, it matches the findings of \cite{Abdo2010c}. One of the reasons this may happen in 3C 273 is the fact that among the luminosities tested, the lowest and highest values (with exception of a single-point) are not present in this analysis (see panel a on Figure~\ref{mw} and the top panel of Figure~\ref{gamind}). This means that the gamma-ray spectrum during the lowest and highest activity states (once again, with the exception of a single-point) is not well-described with a power-law spectral model. Therefore, the lack of a correlation might just be indicative that the gamma-ray luminosity range is not large enough to test for a correlation.\\

\section{Correlation Analysis} \label{correlations} 

It is widely known that in blazars, the millimeter and radio emission are dominated by synchrotron emission from the jet \citep[e.g.,][and references therein]{Rani2011,Georganopoulos2012,Monte2012,Bottacini2016,Bottcher2017,Singh2020}. 

Following \cite{Aleksic2014} and \cite{PatinoAlvarez2018}, we decided to test if the optical emission has the same origin as the millimeter and radio. We took quasi-simultaneous observations (within 24 hours) between the V-band and the 1mm and applied a Spearman Rank Correlation test, to determine if the variations in both bands are related. The same correlation test was performed between the V-band and the 15~GHz light curves. The analysis yielded a correlation coefficient ($\rho$) of 0.05 and a p-value of 0.54 between the V-band and the 1mm emission, meaning that no correlation was found. On the other hand, for the test between the V-band and the 15~GHz light curve, the analysis results in a correlation coefficient ($\rho$) of -0.23 with a p-value of 0.001; while the p-value shows high significance, the correlation coefficient is too low, $\rho^2$=0.05, meaning that only 5\% of the variability in one of the bands, can be related or attributed to changes in the other band. Since the errors in the V-band and in the 15~GHz data are in the order of $\rho^2$ ($\sim$5\%), at best, we can speculate that there is a weak anti-correlation between the V-band and the 15 GHz, which will be discussed in more detail.

\begin{figure}
\includegraphics[width=0.48\textwidth]{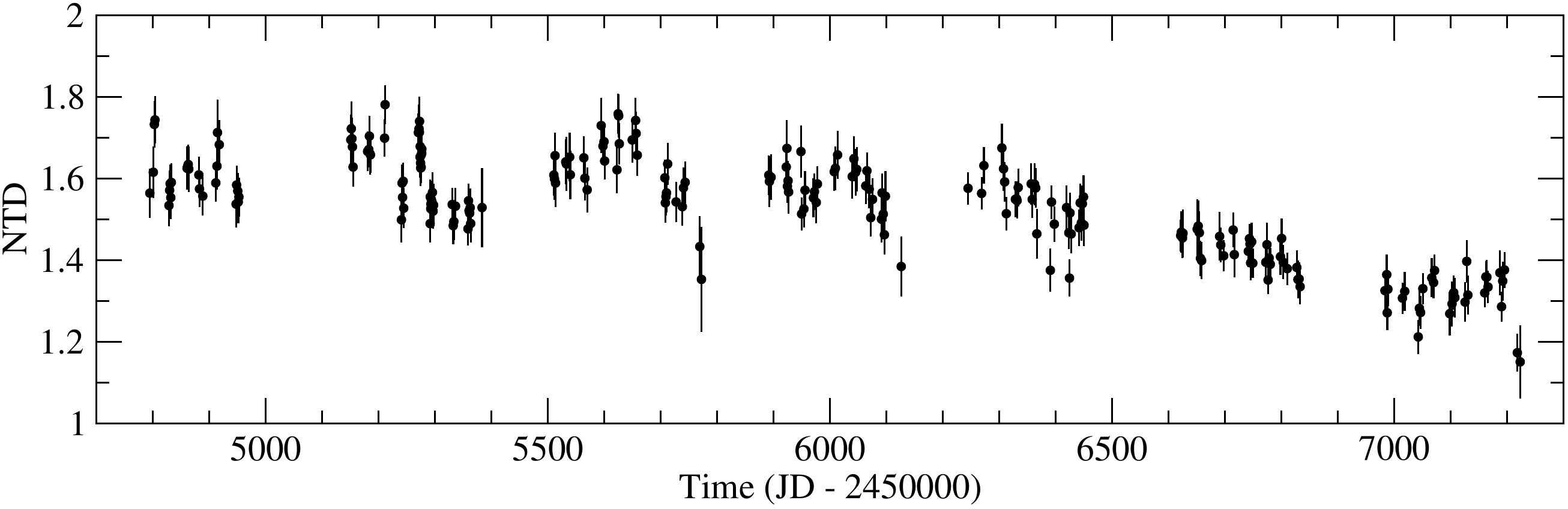}
\caption{Light curve of non-thermal dominance for 3C 273.}
\label{NTD}
\end{figure}

The results from the Spearman correlation analysis, suggest that the optical emission does not share the same synchrotron origin as the 1mm and the 15~GHz, suggesting that the optical emission is not dominated by the jet, but by thermal radiation from the accretion disk. To verify this result, we decided to compute the Non-Thermal Dominance parameter \citep[NTD,][]{Shaw2012,PatinoAlvarez2016,Chavushyan2020} using the H$\beta$ emission line and the 5100 \AA\ continuum luminosities, as well as the luminosity relation presented by \cite{Greene-Ho2005}. As explained in the aforementioned papers, a NTD value of one indicates that the optical emission is produced by thermal processes only, in this case, the accretion disk. A NTD value between one and two indicates the existence of a jet, but with the continuum still being dominated by the accretion disk. A NTD value of two indicates the presence of a jet with a luminosity equal to that of the accretion disk; while a NTD value higher than two means that the jet is the dominant source of emission. For 3C 273, the NTD value is between one and two for the entire time-frame of study, further supporting our hypothesis. The NTD light curve can be found in Figure~\ref{NTD}.

To further test our hypothesis, as well as to quantify any correlations between the behaviour of different bands, we carried out Cross-Correlation analysis using three different methods: The interpolation method \citep[ICCF,][]{Gaskell1986}, the Discrete Cross-Correlation Function \citep[DCCF,][]{Edelson1988} and the Z-Transformed Discrete Correlation Function \citep[ZDCF,][]{Alexander1997}. We used the modified versions of these methods as described in \citet{PatinoAlvarez2013}. To establish a significance to the correlation coefficients we performed Montecarlo simulations following \cite{Emmanoulopoulos2013} and \cite{PatinoAlvarez2018}. The confidence intervals (at 90 \%) in the delay were obtained as in \cite{PatinoAlvarez2018}. To determine the truthfulness of the delays obtained with the cross-correlation analysis, we performed an alias check via Fourier analysis \citep{Press2007}\footnote{\url{https://www.cambridge.org/numericalrecipes}}. The resulting delays obtained from the cross-correlation analysis are summarised in Table \ref{lags1}.\\ 

\begin{figure}
\includegraphics[width=0.5\textwidth]{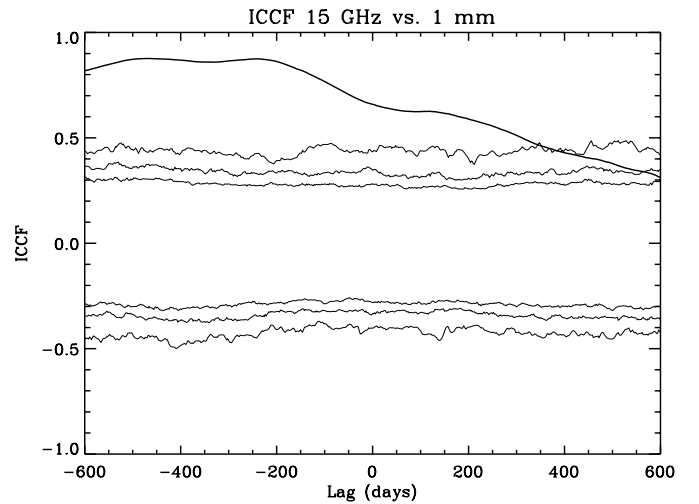}
\caption{Example of inconclusive cross correlation result. The noisy lines represent the significance of correlation at 90, 95, and 99 \%, below and above correlation coefficient zero. The smooth line is the interpolated cross-correlation function (ICCF).}
\label{cc_weird}
\end{figure}

It is important to mention that there are non-conclusive results for some of the light curve pairs, where lag periods from 200 days to 1000 days, appear as significant correlations (see Figure~\ref{cc_weird}). We explain this effect by taking into account the low-amplitude variability over large periods of the light curves. When the cross-correlation coefficient is calculated for every lag, light curves in quiescent states will be very similar to each other, regardless of the time axis-shift. The same issue also contaminates cross-correlation functions for correlated signals.\\

\begin{table*}
\centering
\begin{minipage}{100mm}
\caption{Delays obtained via cross-correlation analysis, for the entire time-range of study. Delays are in units of days. All correlations are at a confidence level $>99\%$, unless otherwise stated. \textbf{The uncertainty is at a 90\% confidence level.}}

\begin{tabular}{ccccc}
\hline
Bands & Delay \\
\hline
\hline
1mm vs. 5100 \AA\        & - - - \\
1mm vs. X-rays             & - - -*  \\
1mm vs. Gamma-rays   & - - -  \\
15 GHz vs. 1mm           & - - -** \\
15 GHz vs. 5100 \AA\   & - - -** \\
15 GHz vs. Gamma-rays & - - - \\
15 GHz vs. V Band          & - - -** \\
15 GHz vs. X-rays.          & - - -** \\
5100 \AA\ vs. Gamma-rays & - - - \\
5100 \AA\ vs. X-rays           & - - -* \\
V band vs. 1mm                  & - - - \\
V band vs. Gamma-rays     & - - - \\
V Band vs. 5100 \AA\         & 6.1$\pm$7.0  \\
V Band vs. X-rays             & 16.5$\pm$12.1 \\
X-rays vs. Gamma-rays    & - - - \\
\hline
\multicolumn{2}{l}{* All delays observed were found to be alias.} \\
\multicolumn{2}{l}{** Inconclusive analysis.} \\
\end{tabular}
\label{lags1}

\end{minipage}
\end{table*}



\subsection{V-band and 15~GHz Anticorrelation} \label{v15GHzsec}

\begin{figure}
\includegraphics[width=0.5\textwidth]{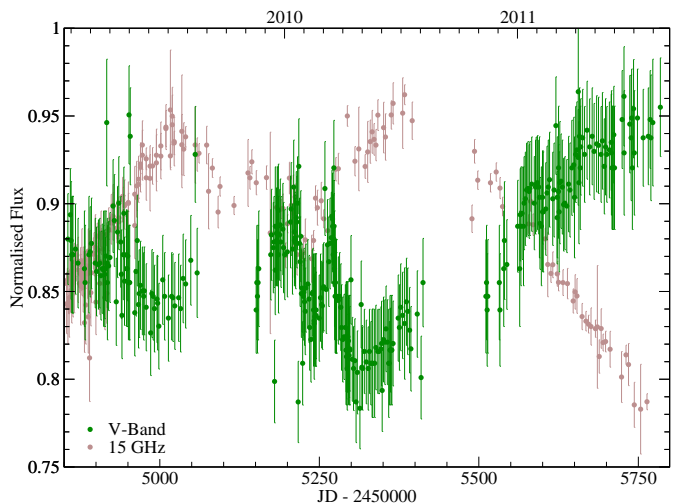}
\caption{Normalized light curves for the V-band and the 15~GHz emission in the period $JD_{245} = 4860 - 5760$.}
\label{v15GHz}
\end{figure}

In Section \ref{correlations} we reported finding a weak anti-correlation between the V-band and 15~GHz. We decided to investigate this further by analysing the normalized light curves, then we noticed that during the time period $JD_{245}=4860 - 5760$, the behaviour of the V-band and the 15~GHz light curves are contrary (see Figure~\ref{v15GHz}). However, this anti-correlation does not hold for the entirety of the time-frame studied in this work; which is the likely reason why it only appeared as a weak anti-correlation on the analysis mentioned above.

\begin{figure}
\includegraphics[width=0.5\textwidth]{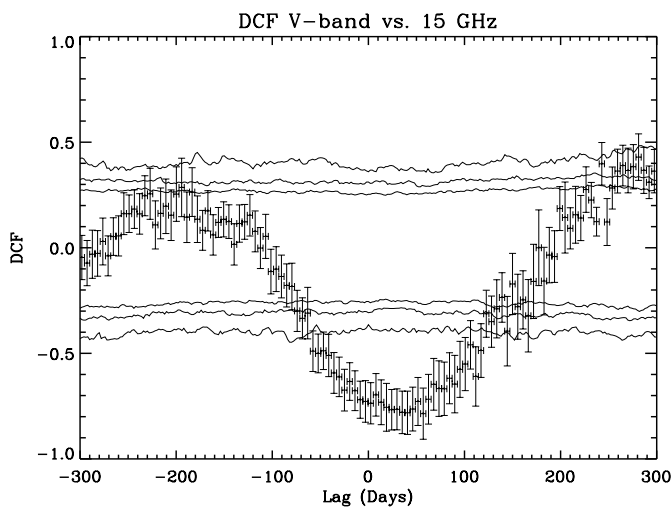}
\caption{Discrete cross-correlation function between light curves for the V-band and the 15~GHz emission during the period $JD_{245} = 4860 - 5760$. The lines represent the significance at 90\%, 95\%, and 99\%, above and below zero. }
\label{v15GHzcorr}
\end{figure}

Moreover, we decided to test it with a cross-correlation analysis, restricted to the period mentioned above. The analysis results in an anti-correlation at a delay of 39.6$\pm$3.2 days, with the V-band leading the 15~GHz emission (see Figure~\ref{v15GHzcorr}). In blazars, the V-band is usually dominated by synchrotron emission from the jet; however, 3C~273 has a history of being dominated by the accretion disk in the optical bands \citep[e.g.][]{Shields1978,Ghisellini2010}. This fact that the optical emission is mostly dominated by thermal emission from the accretion disk (see Section~\ref{correlations}) can shed light into this behaviour, which will be discussed in Section~\ref{results_discussion}.

\subsection{Polarimetric Behaviour} \label{polbehav}

3C~273 is classified as a Low Polarisation Quasar \citep[LPQ,][]{Lister2000}. The degree of polarisation during the time-frame of this study never exceeds 2 \%. In many cases, the variability time-scale of polarization data (both $P$ and {\it PA}) can give hints on the phenomena taking place at the synchrotron emission zone. Due to the ambiguity in {\it PA} \citep[we discuss this ambiguity and its implications in][]{PatinoAlvarez2018} we can only be certain of short-term {\it PA} behaviour in the case of very high cadence data. As can be seen in Figure~\ref{polfig}, the {\it PA} shows trend changes from northward to southward direction, and vice versa during the entire observational time-frame, with angle variations between 30 to 60 degrees. We interpret these swings in the {\it PA} as changes in the magnetic field topology of an emitting region.

For long time-scale trends, we expect the emission region to be large, while for short time scale swings we expect small emitting regions. polarisation swings over short-time scales have been associated with component ejections and shocks moving through the jet and associated flares \citep[e.g.][]{Valtaoja1993, Abdo2010e}; however, there is no constant flaring behaviour to explain constant swings in {\it PA} over the studied time-frame of observations. A possible physical scenario for constant swings in {\it PA} without counterparts in other bands is that the {\it PA} swings are caused by chaotic or tangled magnetic field regions. For the large time-scale trend, it is likely the result of a slowly changing, ordered magnetic field due to an additional linearly rotating component in a large emission region such as the jet itself \citep{Valtaoja1993, Raiteri2013, Covino2015}. Other possibilities suggested in \cite{Raiteri2013} are that the polarised emission originates in a jet with helical magnetic fields or a transverse shock wave moving across the jet. The low degree of polarisation observed in 3C~273, supports the idea that the jet is not dominant in the optical over the studied time-range.


\section{Results and Discussion} \label{results_discussion}

(i) The relationship between the gamma-ray spectral index and luminosity has had multiple relationships published in the literature: harder-when-brighter \citep{Brown2011,FanXL2012,PatinoAlvarez2018}, softer-when-brighter \citep{Foschini2010} and weak/no correlation between spectral index and gamma luminosity \citep{Abdo2010c}. \citeauthor{Abdo2010c} had the largest sample size from the LAT Bright AGN sample \citep{Abdo2009}. However, the observed differing relationships imply that the relation could be object and/or activity state dependent. Indeed, there have been observations of harder-when-brighter during periods of flaring activity\citep{Abdo2010a}. In the case of individual sources, \cite{PatinoAlvarez2018} reported that the gamma-ray spectrum becomes harder with higher luminosity for the source 3C 279, while \cite{Vercellone2010} reports the same trend for 3C 454.3, both FSRQ. In contrast, for 3C 273 we found no significant correlation between the gamma-ray spectral index and the luminosity. As mentioned in Section~\ref{grayspecindex}, this result might be due to sampling problems, however, in the case that the lack of correlation is real, this might mean that the energetics of the gamma-ray production processes causing the variability in 3C 273 are different from other blazars, with changes in the Lorentz factor of the jet, or hadronic processes like proton synchrotron \citep[e.g.,][]{Kundu2014,Esposito2015,Petropoulou2015}.

(ii) We found that the dominant component of the optical emission in 3C 273 does not share the synchrotron origin that the 1mm and the 15~GHz emission have. We did this by applying a Spearman rank correlation test to quasi-simultaneous observations (within 24 hours) of the V-band and the 1mm light curves, as well as the 15 GHz emission. The lack of correlation points towards the optical emission being dominated by thermal emission from the accretion disk during the studied period. The non-thermal dominance values obtained for this source, between one and two, also support the scenario that the thermal emission from the accretion disk dominates the optical emission. Furthermore, the optical polarisation of 3C~273 remains at a very low level, with a maximum of $\sim$ 2 \%. The low polarisation additionally suggests that the jet is not the dominant source of optical emission, as opposed to an object such as 3C~279, where there is a very high level of polarisation and variability that suggests jet dominance of the emission \citep[e.g.][and references therein]{PatinoAlvarez2018}. LPQ's such as 3C~273 are thought to be in quiescent phases where the jet emission is swamped by accretion disk emission \citep[e.g.,][]{Lister2000}. An interesting example of a blazar being dominated by thermal or non-thermal emission depending on the activity state of the source is CTA 102, where the optical emission is dominated by the accretion disk during quiescent states, and dominated by the jet during flaring states \citep{Chavushyan2020}.

(iii) We report the finding of an anti-correlation between the V-band and the 15~GHz light curves during the time period $JD_{245}=4860 - 5760$, at a delay of 39.6$\pm$3.2 days (see Section~\ref{v15GHzsec}). To explain this behaviour, we have to understand what causes both bands to both drop and increase, and how those processes can be related. A scenario that fits the observed behaviour, is the case where the inner part of the accretion disk falls into the black hole, which causes a drop in the X-ray emission; this event is normally followed by the ejection of a component from the jet base \citep[e.g.,][]{Marscher2002,Chatterjee2009,Arshakian2010,Casadio2015}. Given that the X-ray and optical emission in this source are tightly correlated (see Table~\ref{lags1}), it is feasible that such an event will cause a decrease in the optical emission, while after some time, causing an increase in the radio jet emission due to the ejection of a new component.


\section{Conclusion} \label{conc}

We have presented results of a multiwavelength analysis of the blazar 3C~273. Our main findings are summarised as follows.
\begin{itemize}
\item We find that thermal emission from the accretion disk dominates the optical emission in this object over the studied time range. This is supported by the low level of polarisation, the lack of a strong correlation between the optical and the 1mm and 15~GHz bands, as well as the non-thermal dominance values obtained.  \\
\item We found no correlation between the gamma-ray luminosity and the gamma-ray spectral index, however, we cannot confirm if it is due to a sampling bias or if the lack of correlation is real.\\
\item We find an anti-correlation of V-band and 15~GHz during the time period of  $JD_{245}=4860 - 5760$, which we explain with the argument that an ejection from the jet, and therefore an increment in radio synchrotron emission, is caused after the inner part of the accretion disk falls into the black hole, therefore causing a drop in the accretion disk emission.\\
\end{itemize}


\section*{Acknowledgments}

The authors thank the anonymous referee for their helpful comments that led to an improvement of this article. We also wish to thank P. Smith for his invaluable assistance with the polarimetry analysis. This work was supported by CONACyT (Consejo Nacional de Ciencia y Tecnolog\'ia) research grants 151494 and 280789 (M\'exico)  S.F. acknowledges support from the University of Texas at San Antonio (UTSA) and the Vaughan family, support from the NSF grant 0904421, as well as the UTSA Mexico Center Research Fellowship funded by the Carlos and Malu Alvarez Fund. V. M. P.-A. acknowledges support from the CONACyT program for Ph.D. studies. V. M. P.-A. acknowledges support from the Max Planck Society via a Max Planck Partner Group led by V. M. P.-A. This work has received computational support from Computational System Biology Core, funded by the National Institute on Minority Health and Health Disparities (G12MD007591) from the National Institutes of Health. Data from the Steward Observatory spectropolarimetric monitoring project were used; this program is supported by Fermi Guest Investigator grants NNX08AW56G, NNX09AU10G, and NNX12AO93G. 1~mm flux density light curve data from the Submillimeter Array was provided by Mark A. Gurwell. The Submillimeter Array is a joint project between the Smithsonian Astrophysical Observatory and the Academia Sinica Institute of Astronomy and Astrophysics and is funded by the Smithsonian Institution and the Academia Sinica. This research has made use of archival data from the OVRO 40-m monitoring program \citep{Richards2011} which is supported in part by NASA grants NNX08AW31G and NNX11A043G, and NSF grants AST-0808050 and AST-1109911. \\

\section*{Data Availability}

The data underlying this article will be shared on reasonable request to the corresponding author. \\


\bibliographystyle{mnras} 

\bibliography{3C273_refs} 



%
%


\bsp	
\label{lastpage}
\end{document}